\documentclass[british]{aa}  
\usepackage[british]{babel}           % ngerman: neue deutsche Rechtschreibung
\usepackage{graphicx}                 
\usepackage{subfig}
\usepackage{caption}
\usepackage{everysel}
\usepackage{keyval}
\usepackage{ragged2e}

\newcommand{\xmm}{XMM-Newton}
\newcommand{\rxj}{RX\,J0720.4$-$3125}
\begin{document}
   \title{Spectral and temporal variations of the isolated neutron star RX\,J0720.4$-$3125: 
          new XMM-Newton observations
	  \thanks{Based on observations with XMM-Newton, an ESA Science Mission with 
	          instruments and contributions directly funded by ESA Member states and the USA (NASA)}}
   
\titlerunning{Spectral and temporal variations of \rxj}
   \subtitle{}

   \author{M.M. Hohle\inst{1,2} \and F. Haberl\inst{1} \and J. Vink\inst{3}                                                                                     \and R. Turolla\inst{4,5}
                                \and V. Hambaryan\inst{2}
                                \and S. Zane\inst{5}
                                \and C.P. de Vries\inst{6}
                                \and M. M\'endez\inst{7,8}
%                               \and F. Verbunt\inst{}
          }
\authorrunning{Hohle et al.}
   \offprints{Markus Hohle, mhohle@astro.uni-jena.de}

   \institute{Max-Planck-Institut f\"ur extraterrestrische Physik, Giessenbachstrasse, 85741 Garching, Germany
\and Astrophysikalisches Institut und Universit\"ats-Sternwarte Jena, Schillerg\"asschen 2-3, 07745 Jena, Germany 
\and University Utrecht, PO Box 80000, 3508 TA Utrecht, The Netherlands
\and Department of Physics, University of Padua, via Marzolo 8, 35131 Padova, Italy 
\and Mullard Space Science Laboratory, University College London, Holmbury St. Mary, Dorking, Surrey, RH5, 6NT, UK
\and SRON, Netherlands Institute of Space Research, Sorbonnelaan 2, 3584 CA, Utrecht, The Netherlands
\and Kapteyn Astronomical Institute, University of Groningen, P.O. Box 800, 9700 AV Groningen, The Netherlands
\and Astronomical Institute Anton Pannekoek, University of Amsterdam, Kruislaan 403, 1098 SJ Amsterdam, The Netherlands
}
          %    \email{mhohle@astro.uni-jena.de, rne@astro.uni-jena.de}
        % \and
          %   University of Alexandria, Department of Geography, ...\\
           %  \email{rne@astro.uni-jena.de}
           %  \thanks{The university of heaven temporarily does not
           %          accept e-mails}

 %  \date{Received September 15, 1996; accepted March 16, 1997}

% \abstract{}{}{}{}{} 
% 5 {} token are mandatory
 
  \abstract
{In the past, the isolated, radio-quiet neutron star \rxj\ showed variations in its spectral parameters (apparent radius, temperature of the emitting area and equivalent width of the absorption feature) seen in the X-ray spectra, not only during the spin period of 8.39~s, but also over 
time scales of years. New X-ray observations of \rxj\ with \xmm\ extend the coverage to about 7.5 years with the latest pointing performed in November 2007. Out of a total of fourteen available EPIC-pn datasets, eleven
have been obtained with an identical instrumental setup (full
frame read-out mode with a thin filter), and are best suited for
a comparative investigation of the spectral and timing properties
of this enigmatic X-ray pulsar. }
%aims
{We analysed the new \xmm\ observations together with archival data in order to follow the 
spectral and temporal evolution of \rxj.}
%methods
{All \xmm\ data were reduced with the standard XMM-SAS software package. A systematic and consistent data reduction of all 
these observations is warranted in order to reduce systematic errors as far as possible.}
%Results
{We investigate the phase residuals derived from data from different energy bands using different timing solutions for the spin
period evolution and confirm the phase lag between hard and soft photons. The phase shift in the X-ray pulses between hard and soft 
photons varies with time and changes sign around MJD=52800 days, regardless of the chosen timing solution. The phase residuals
show a marked dependence on the energy band, and possibly follow a cyclic pattern with a period of $\sim$9-12 yrs. We find that an abs(sine) dependence provides a better fit to the residuals with respect to a simple sine wave, although in both cases the reduced $\chi^2$ is not completely satisfactory. 
We compared the model fit to phase residuals derived from EPIC-MOS and Chandra (HRC and ACIS) 
data restricted to the hard energy band (400-1000 eV) to take into account the different energy responses of these instruments.}
%conclusions
{The new data are not in contradiction with \rxj\ being a precessing neutron star but do not provide overwhelming evidence for this picture either. 
}


\maketitle
%
%________________________________________________________________

\section{Introduction}
The radio quiet, isolated neutron star \rxj\ was discovered in ROSAT all-sky survey data (\cite{1997A&A...326..662H}).
It belongs to a group of seven neutron stars with similar properties often called the magnificent seven, hereafter M7. 
The X-ray spectra are well represented by a blackbody model (with absorption features in some cases) without any additional 
non-thermal component. 
For the two brightest  stars (RX J1856.5-3754 and \rxj) the trigonometric parallax was measured 
(Kaplan et al., \cite{Kaplandist1856}) which enables us to estimate the size of the emission region. 
Six of them are X-ray pulsars with relatively long periods between 3.5~s and 11.4~s (for a review see, e.g., Haberl, \cite{Hab07}). 
From the spin period, period derivative ($\dot{P}$) and the energy of absorption lines one can independently estimate the magnetic 
field strength on the surface of the neutron star. Both methods consistently suggest strong magnetic fields in excess of 
$\sim10^{13}$ Gauss (Haberl, \cite{Hab05}). The properties of the M7 make them of special interest for the investigation 
of cooling models of neutron stars (see Page \& Reddy, \cite{page1}; Page et al. \cite{page2} and Blaschke \& Grigorian, \cite{bug}) 
and for constraining of the equation of state of matter at nuclear densities (Tr\"umper et al., \cite{trümper}).

\rxj\ shows a long term variation in its spectral parameters (see de Vries et al., \cite{Vries04}; Vink et al., 
\cite{Vinkvar} and Haberl et al., \cite{Hab06}, hereafter H06), i.e. its temperature, size of the emitting area and equivalent width of 
the absorption feature. The variations derived from data covering 5.5 years were consistent with a period of about 
(7.1 $\pm$ 0.5)~yr, when assuming a sinusoidal shape of the variations (H06). 
Kaplan \& van Kerkwijk (\cite{KuK05}) (hereafter KvK05) performed a coherent timing analysis for \rxj\ by including 
all data (ROSAT, \xmm\ and Chandra) available at that time. This enabled them to determine the period more accurately 
then previous studies (Cropper et al., \cite{cropper1}, Zane et al., \cite{zane} and Kaplan et al., \cite{KuK02}). In their Table 2 KvK05 
provide two different values for a constant $\dot{P}$: $7.016\times 10^{-14}$~s/s for Chandra data only and 
$6.983\times 10^{-14}$~s/s using all data. By applying these timing solutions to derive phase binned light curves, KvK05 
obtained phase residuals. These phase residuals, as shown by H06, can be modeled by a sinusoidal function with a period of 
(7.7 $\pm$ 0.6)~yr. This is consistent with the (7.1 $\pm$ 0.5) yr period of the spectral variations. 
As a possible explanation for the behaviour of \rxj, de Vries et al. (\cite{Vries04}) and H06 suggested precession of the neutron star. van Kerkwijk et al. (\cite{KuK07}, hereafter vK07) 
derived a new $\dot{P}=6.957\times 10^{-14}$~s/s including new \xmm\ and Chandra data, but still obtained sinusoidal-like phase residuals. KvK05 and vK07 derived $\chi^2$=1 for their timing solutions only after adding an artificial uncertainty to the time of arrivals (see KvK05 and vK07 for more details).
Furthermore, they add a glitch model to the spin-down term and claim the occurrence of the glitch around MJD=52800~days 
to explain the timing and spectral changes of this neutron star. Equivalent to the glitch model vK07 found a period $\approx$4.3~yr but that is not consistent with the period $\approx$7.5~yr found by H06. On the other hand, the data set used in H06 (EPIC-pn) is quite different to that used in vK07 (also including EPIC-MOS1 \& MOS2 and Chandra HRC-S/LETG).
 
This paper includes new data from the recent \xmm\ (Jansen et al., \cite{jansen}) observations which are part of an ongoing monitoring programme, and Chandra archival 
data. We discuss our results as new evidence for a cyclic behaviour of \rxj\ based on timing solutions from KvK05 and vK07.

%__________________________________________________________________

\section{X-ray observations and timing analysis}

Since the work of Haberl (\cite{Hab07}) was published, four new \xmm\ observations have been performed 
between May 2006 and November 2007 at six month intervals (satellite revolutions 1181, 1265, 1356 
and 1454). All new EPIC-pn (Str\"uder et al. \cite{strueder}) observations were executed with the 
same instrument setup in full frame mode and with thin filter (see Table~\ref{xmm-obs} for a summery of the observations). Inclusion of the latest pointings brings the number of available EPIC-pn datasets to fourteen. These, in conjunction with the MOS (Turner et al., \cite{turner}) data (see Table~\ref{xmm-obs} for details on the instrumental setup), have been used for the source timing analysis. All observations 
were reduced with the \xmm\ Science Analysis System (SAS) version 7.1.0. 

To derive the phase residuals from a certain timing solution, we binned the light curve in 40, 100 and 400 phase bins. The 
number of phase bins did not influence the result. Considering the time resolution of the full frame (2.6 s) and small 
window mode (0.3 s) we divide the MOS light curves in 10 and 40 phase bins, respectively. Phase residuals from all observations 
are obtained fitting a sinusoidal function to the pulse profile. The uncertainties of the phase residuals are derived from 
the uncertainties of the sinusoidal fit. All errors on the phase residuals in this paper correspond to 1$\sigma$.

In the analysis of KvK05 the phase residuals for the last observations show an increasing trend while in the case of vK07 (with more data included) the trend is reversed. Therefore, the timing solution for \rxj\ strongly depends on the behaviour of the source in the future and all solutions obtained so far must be regarded as preliminary. Hence, in the following we examine phase residuals for two timing solutions provided by KvK05 and vK07 (in both cases their "all data solution").

\begin{figure}
  \centering
   \resizebox{\hsize}{!}
{
   \includegraphics[viewport=85 254 490 572, width=0.48\textwidth]{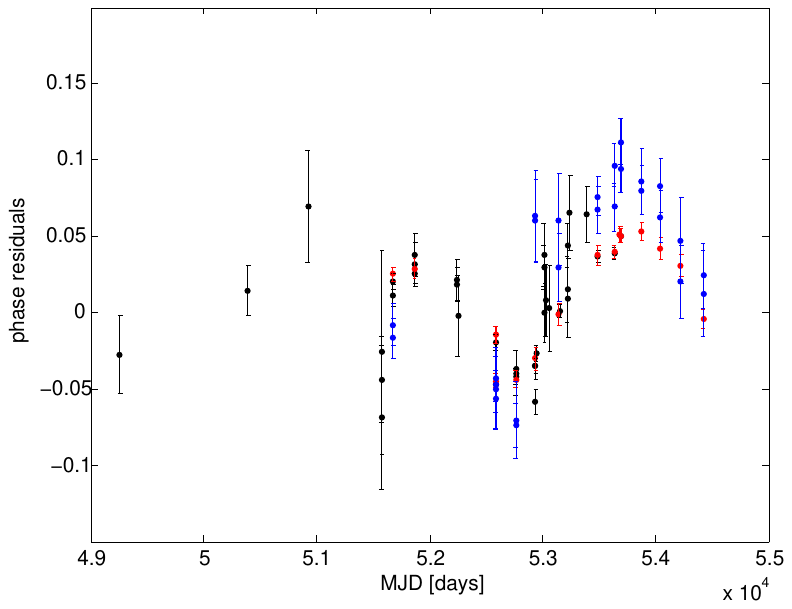}
}
   \caption{Phase residuals of \rxj\ derived from the timing solution from Kaplan \& van Kerkwijk (\cite{KuK05}). 
            Red dots: refer to EPIC-pn data, blue dots: to
	    EPIC-MOS1 \& MOS2 observations, black dots: reproduced from the original solution of Kaplan \& van Kerkwijk (\cite{KuK05}).}
              \label{phaseresall}
\end{figure}

\begin{figure}
  \centering
   \resizebox{\hsize}{!}
{
 \includegraphics[viewport=20 170 565 605, width=0.48\textwidth]{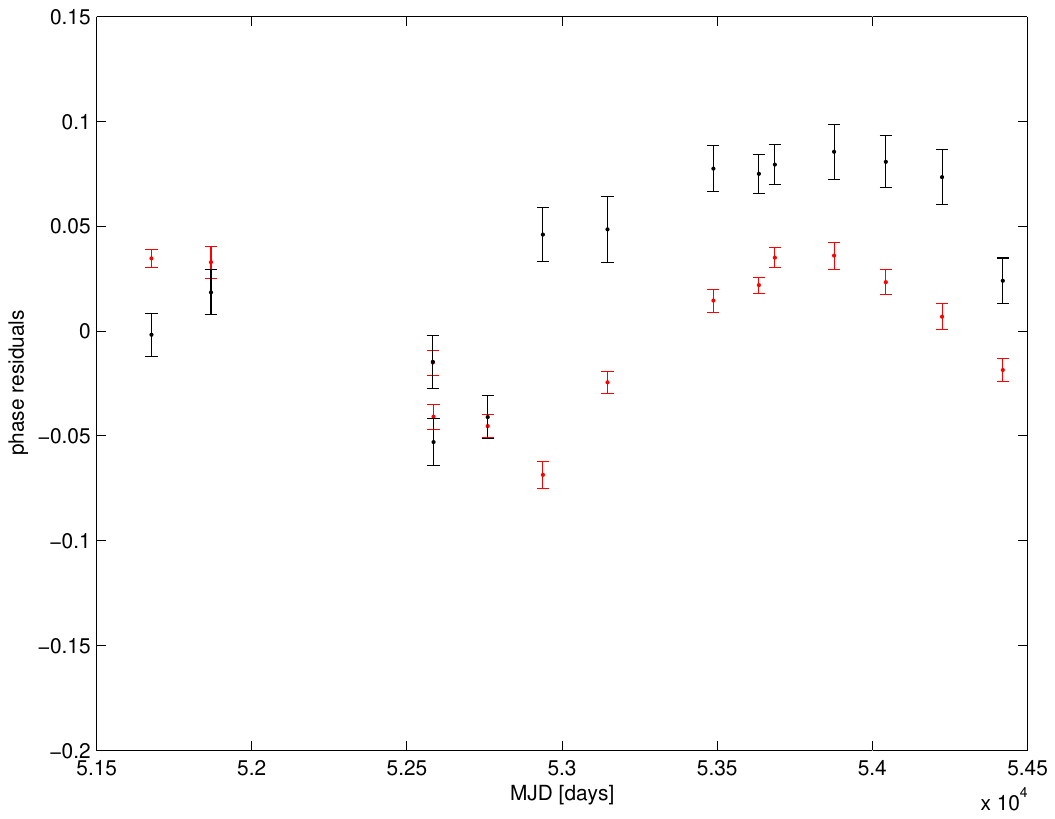}
}
   \caption{Phase residuals of \rxj\ derived from EPIC-pn data with the timing solution from Kaplan \& van Kerkwijk (\cite{KuK05}) 
            in the soft band (120 - 400 eV, red) and hard band (400 - 1000 eV, black).}
              \label{phaseres-redsoftblackhard_KuK05}%
\end{figure}

\begin{figure}
  \centering
   \resizebox{\hsize}{!}
{
  \includegraphics[viewport=85 254 490 572, width=0.48\textwidth]{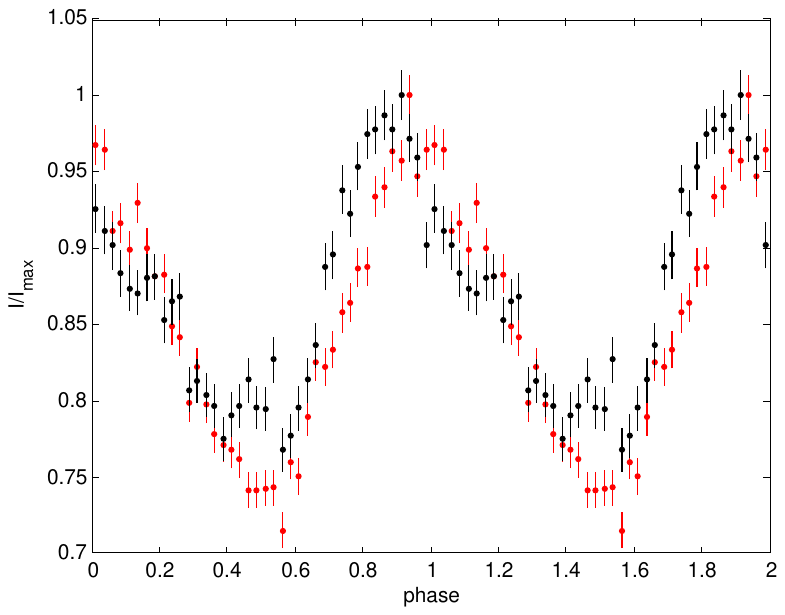}
}
   \caption{Phase shift of hard (black) and soft (red) band of \rxj\ derived from EPIC-pn data (satellite orbit 986) with the timing solution from Kaplan \& van Kerkwijk (\cite{KuK05}). Errors are Poissonian. The y-axis denotes the relative counts per phase bin.}
          \label{lightcurves0986}%
\end{figure}

%Extending the work of Haberl (\cite{Hab07}) we included the results of the timing analysis of the new EPIC-pn observations 
%and also  the EPIC-MOS data in small window and full frame read-out mode. 
To reduce systematic effects due to the changing low-energy 
response of the EPIC-MOS cameras we use only events from the 400-1000 eV band for MOS (the 120-1000 eV band is used for pn).
The phase residuals using the coherent timing solution of KvK05 are shown in Fig.~\ref{phaseresall}. 
As already mentioned in Haberl (\cite{Hab07}), systematic differences between phase residuals from different instruments can 
be seen, which are probably caused by the energy dependent pulse profile. To verify this, we investigated the phase residuals 
for two different bands (soft: 120-400 eV and hard: 400-1000 eV) using EPIC-pn data, and the results are shown in 
Fig. \ref{phaseres-redsoftblackhard_KuK05}. The phase residuals exhibit 
a clear energy dependence while the phase shift between the two bands varies with time and changes sign around MJD=52800 days. As an example of a relatively large phase shift between hard and soft bands we show the pulse profile for satellite orbit 986 in Fig. \ref{lightcurves0986}. The phase shift is clearly visible and the trend is much the same as in Fig. \ref{phaseres-redsoftblackhard_KuK05} if we add or subtract the error of $\dot{P}$ provided in KvK05.

\begin{figure}
  \centering
   \resizebox{\hsize}{!}
{
  \includegraphics[viewport=85 254 490 572, width=0.48\textwidth]{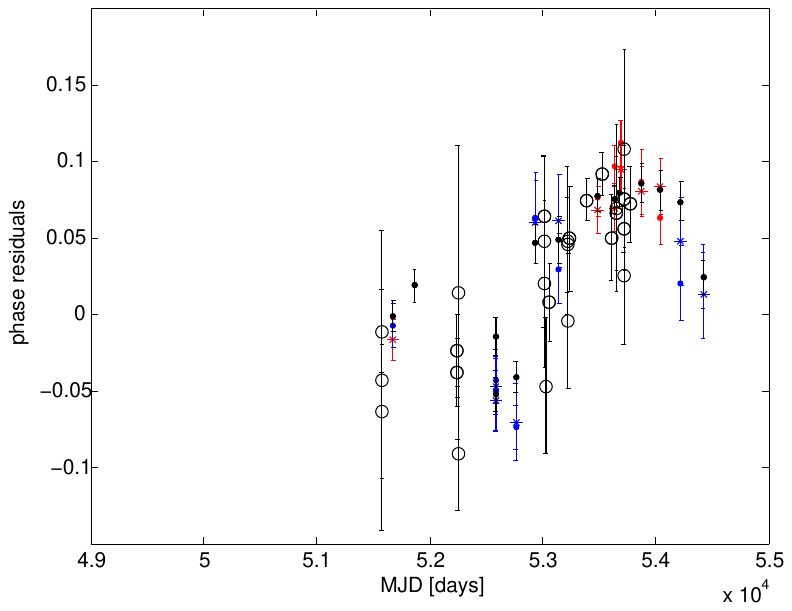}
}
   \caption{Phase residuals of \rxj\ derived from the timing solution from Kaplan \& van Kerkwijk (\cite{KuK05}). 
            Black dots: EPIC-pn, red dots: EPIC-MOS1 small window and thin filter, red stars: EPIC-MOS2 small window and thin 
	    filter, blue dots: EPIC-MOS1 full frame and thin filter, blue stars: EPIC-MOS2 full frame and thin filter (all in hard band); open circles show phase residuals from Chandra HRC (no energy selection) and ACIS (hard band).}
              \label{allphaseres_chandra_all}%
    \end{figure}

We also applied the timing solution from KvK05 to archival Chandra HRC-S/LETG (zeroth order) and ACIS data (for details of the 
observations see KvK05 and vK07). For observations 368, 369, 745, 2771, 2772, 4667, 4668, 
4669, 4670, 4671, 4672, 7177, 7243 and 7245 we use 10 phase bins and 40 phase bins for the other observations 
because of the different time resolution and number of photons. Considering the energy dependence of the phase residuals we 
restrict the energy band for the analysis of the ACIS data to the hard band (400-1000 eV). 
We show all phase residuals from Chandra and \xmm\ observations in Fig.~\ref{allphaseres_chandra_all}.
Comparison with Fig.~\ref{phaseresall} shows that the agreement between the different instruments is much improved.

\begin{figure*}
 \captionsetup[subfloat]{labelformat=empty}
 \centering
 \subfloat[]{\includegraphics[width=0.48\textwidth]{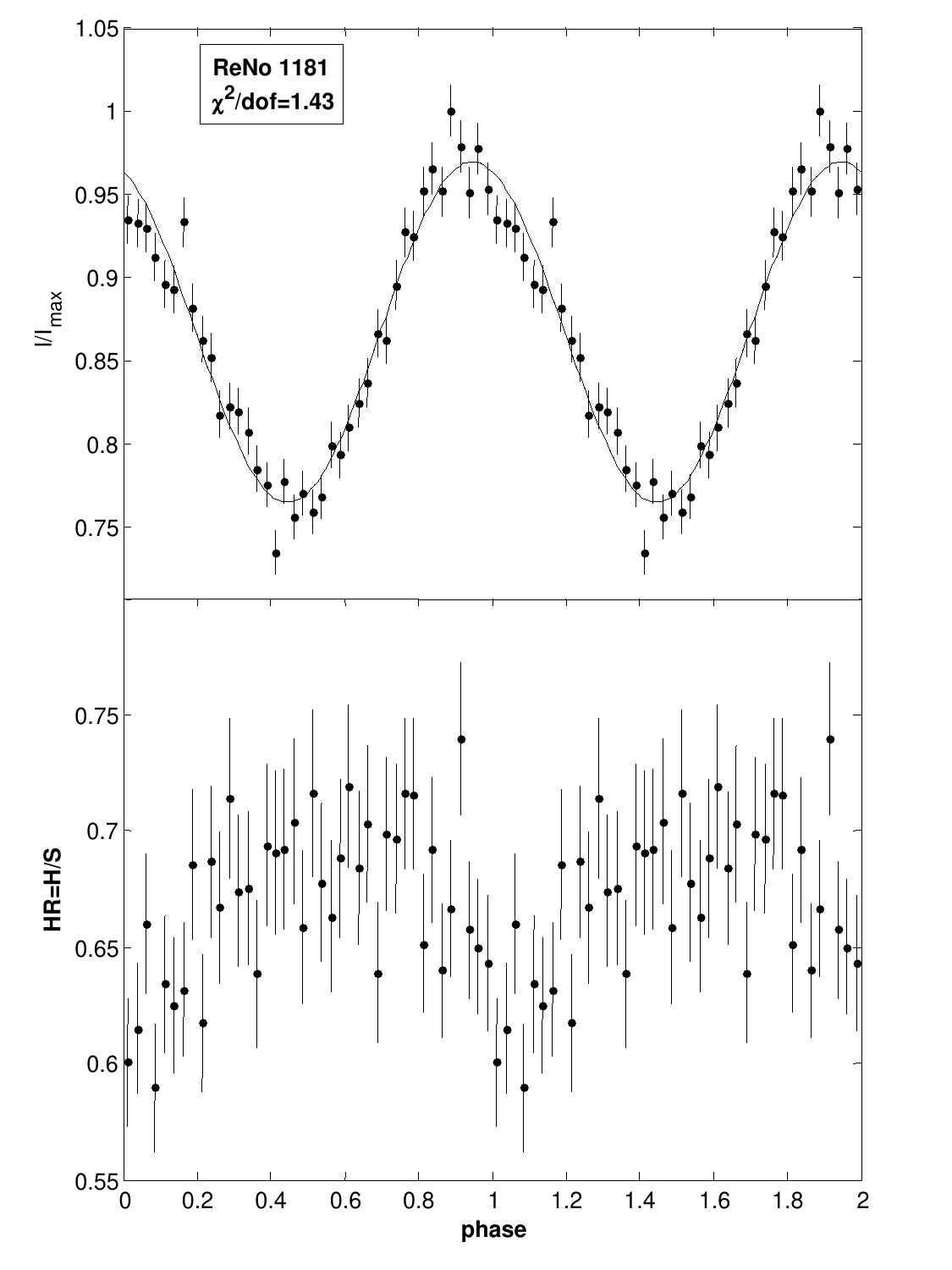}
 \label{fig:1181}}
 \subfloat[]{\includegraphics[width=0.48\textwidth]{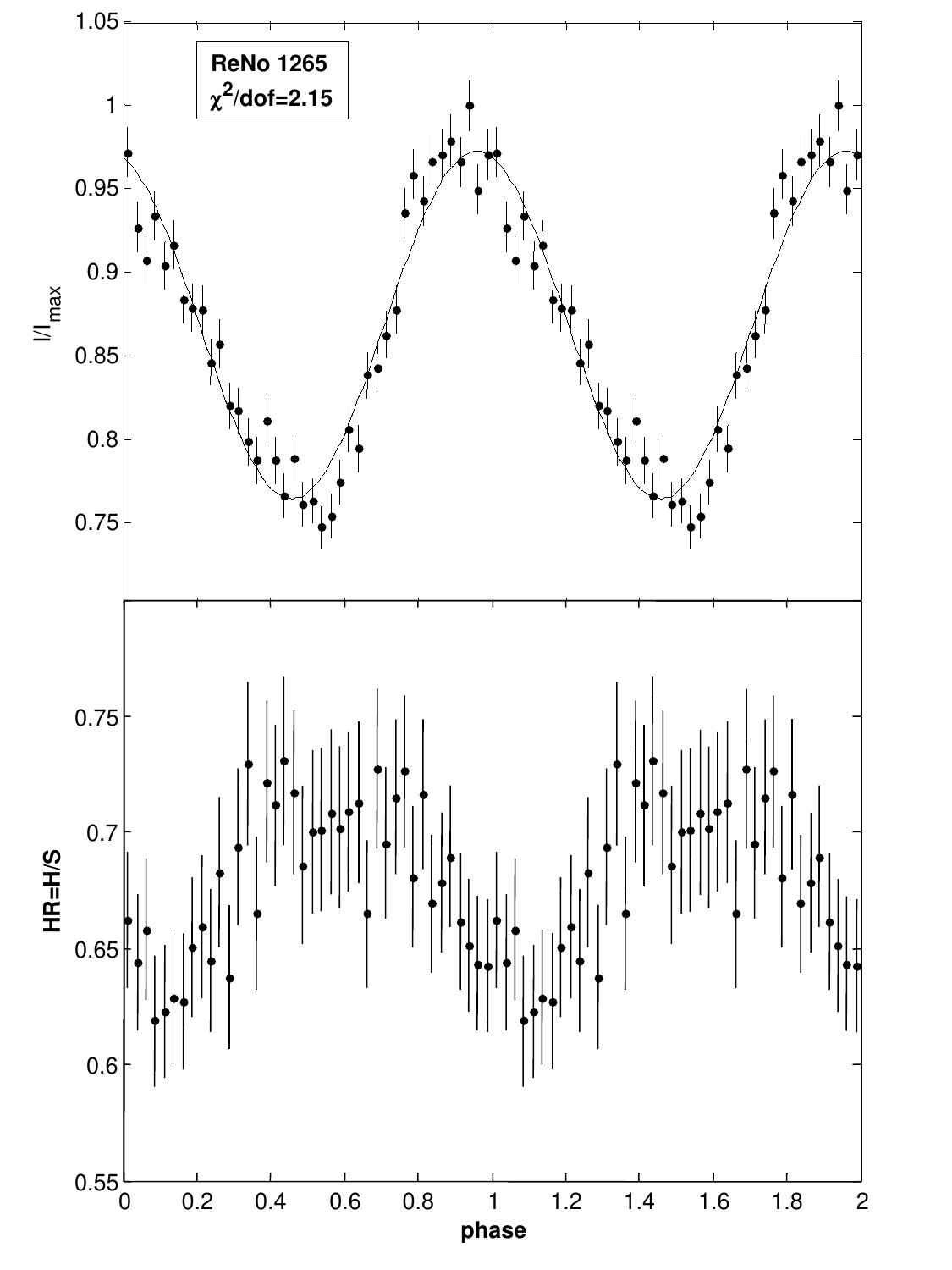}
 \label{fig:1265}}\\\vspace{-1.5em}
 \subfloat[]{\includegraphics[width=0.48\textwidth]{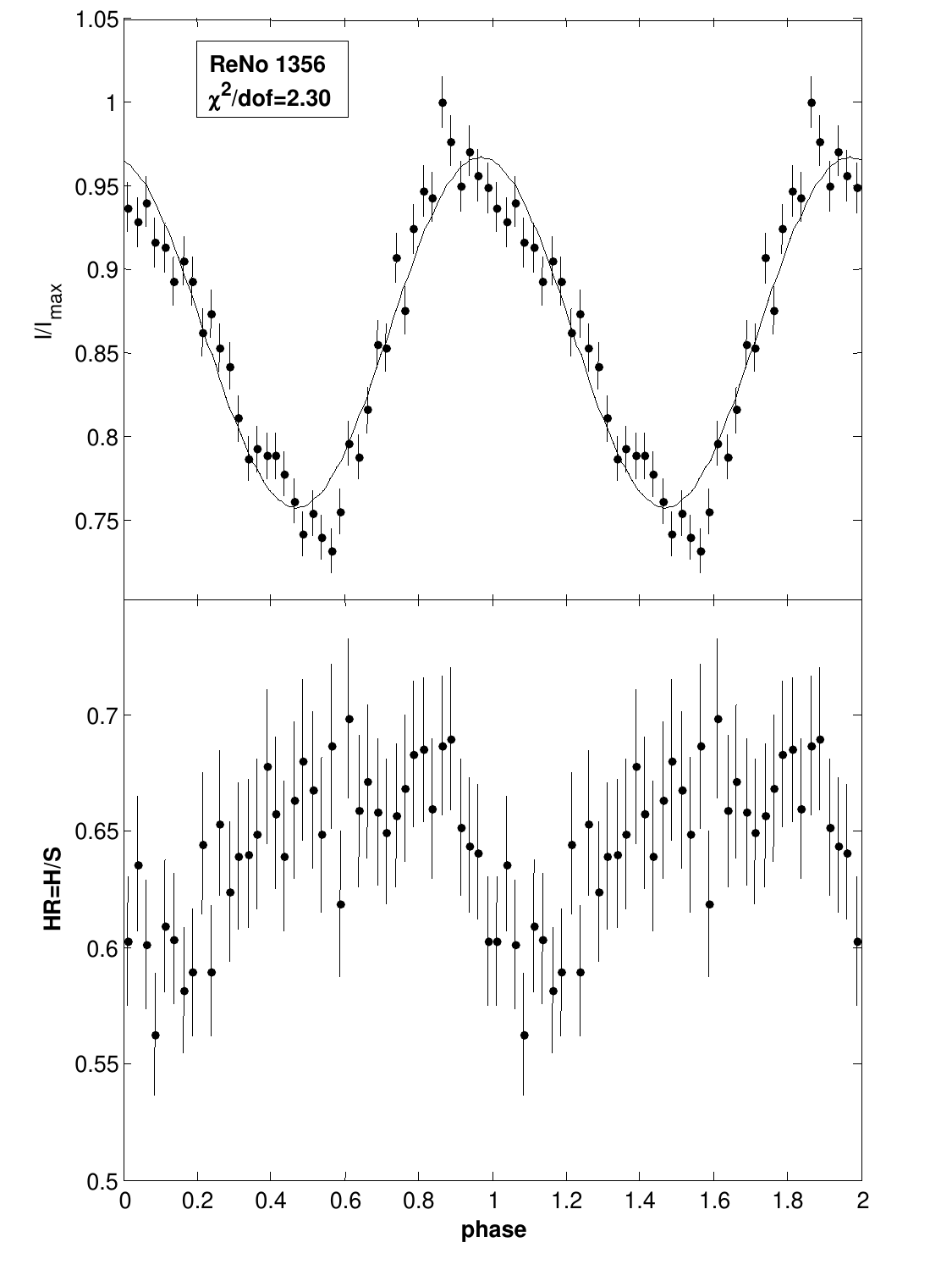}
 \label{fig:1181}}
 \subfloat[]{\includegraphics[width=0.48\textwidth]{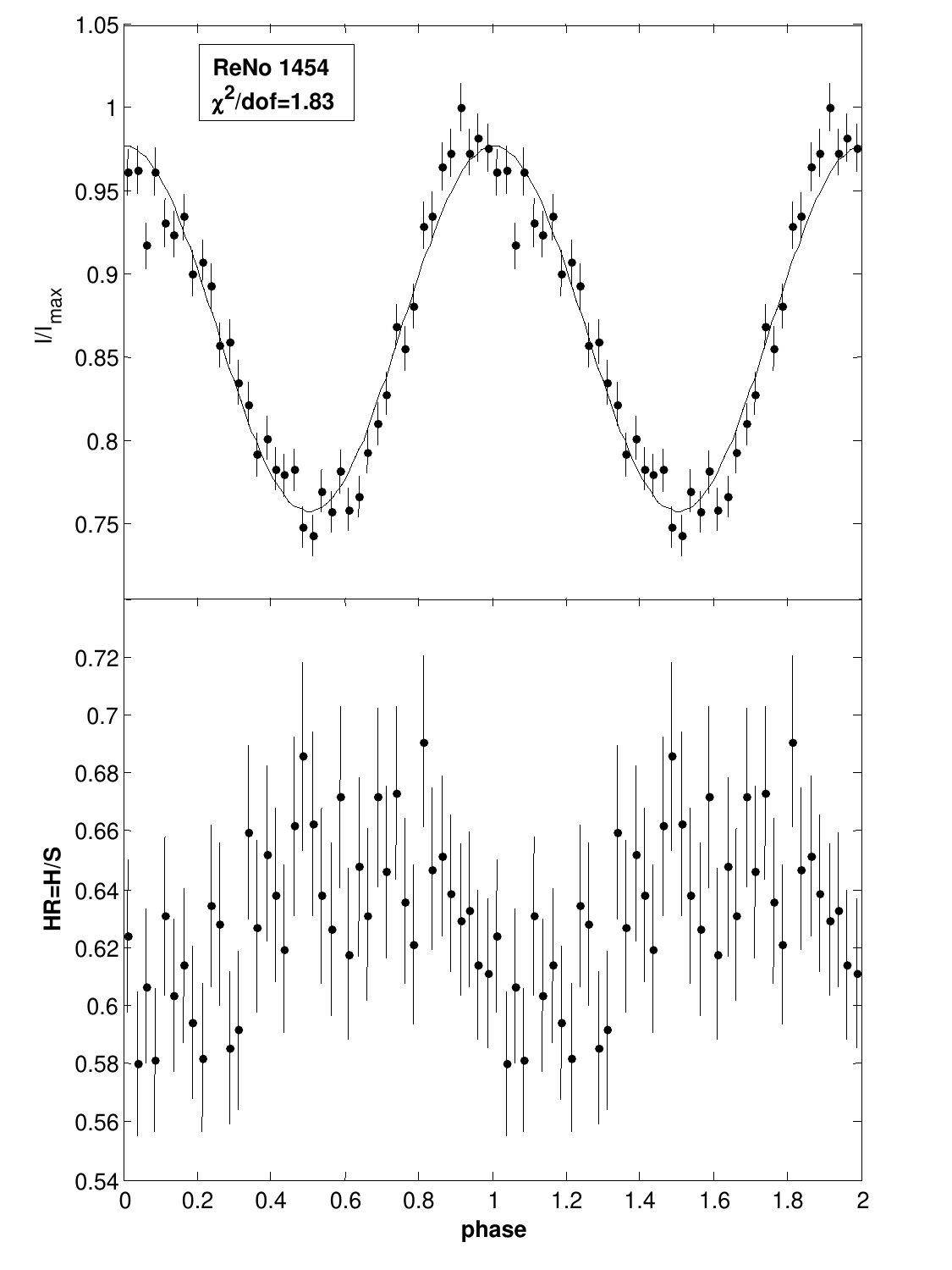}
 \label{fig:1265}}\\\vspace{-1.5em}
 \caption{Folded EPIC-pn pulse profile (120-1000 eV, 40 phase bins, $d.o.f=37$ in all cases) with sinusoidal fit and hardness ratio HR=H/S (bands described in the text) of the last four observations. Pulse phases are obtained by applying the "all data" solution of Kaplan \& van Kerkwijk (\cite{KuK05}). Errors are Poissonian. $I/I_{max}$ denotes the relative counts per phase bin.}
 \label{lightcurves}
\end{figure*}

\begin{figure}
  \centering
   \resizebox{\hsize}{!}
{
   \includegraphics[viewport=85 254 490 572, width=0.48\textwidth]{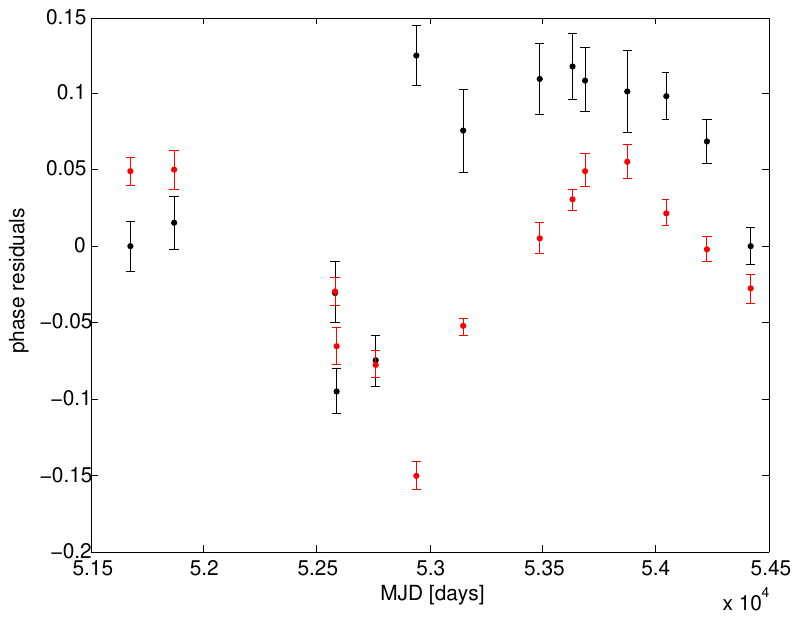}
}
   \caption{Phase residuals of \rxj\ derived from the timing solution from Kaplan \& van Kerkwijk (\cite{KuK05}), 
            similar to those in Fig. \ref{phaseres-redsoftblackhard_KuK05}, but fitting a sawtooth instead of a sinusoidal to the pulse profile.}
              \label{phaseres_sawtooth_KK05}
\end{figure}

As in the EPIC-pn observations before (see Haberl, \cite{Hab07}; H06 and de~Vries~et~al., \cite{Vries04}), the pulse profile can be approximated by a sinusoid. In Fig. \ref{lightcurves} we show the pulse profiles for the last four EPIC$-$pn data sets together with the hardness ratios. The hardness ratios $HR=H/S$ are anti-correlated with intensity. The pulse phases were derived by applying the "all data" solution of KvK05 (\cite{KuK05}). The light curves look the same, except for absolute phase of course, using the timing solution of vK07. As noted before, the profiles are not exactly sinusoidal and look somewhat skewed. This is particularly true for the latest observations. To explore how much the behaviour of the residuals is influenced by the choice of the pulse shape we also tried a sawtooth profile, in which the location and the height of the minimum are free to vary. Results for the phase residuals in the two bands (see Fig. \ref{phaseres_sawtooth_KK05}) are similar to those derived for a sinusoidal profile. This supports the fact that energy-dependent phase residuals are indeed present and do not depend on the chosen shape of the pulse profile. We note that the $\chi^{2}/d.o.f$ values of the sine fits shown in Fig. \ref{lightcurves} are smaller than those of the sawtooth fits ($\chi^{2}/d.o.f=$ 2.94, 4.06, 3.70, 2.50 in the order of increasing revolution number) so that a sine wave provides a better interpretation of the data, although the quality of the fits is still not satisfactory. For all fits $d.o.f=37$.  

\begin{figure}
  \centering
   \resizebox{\hsize}{!}
{
   \includegraphics[width=0.5\textwidth]{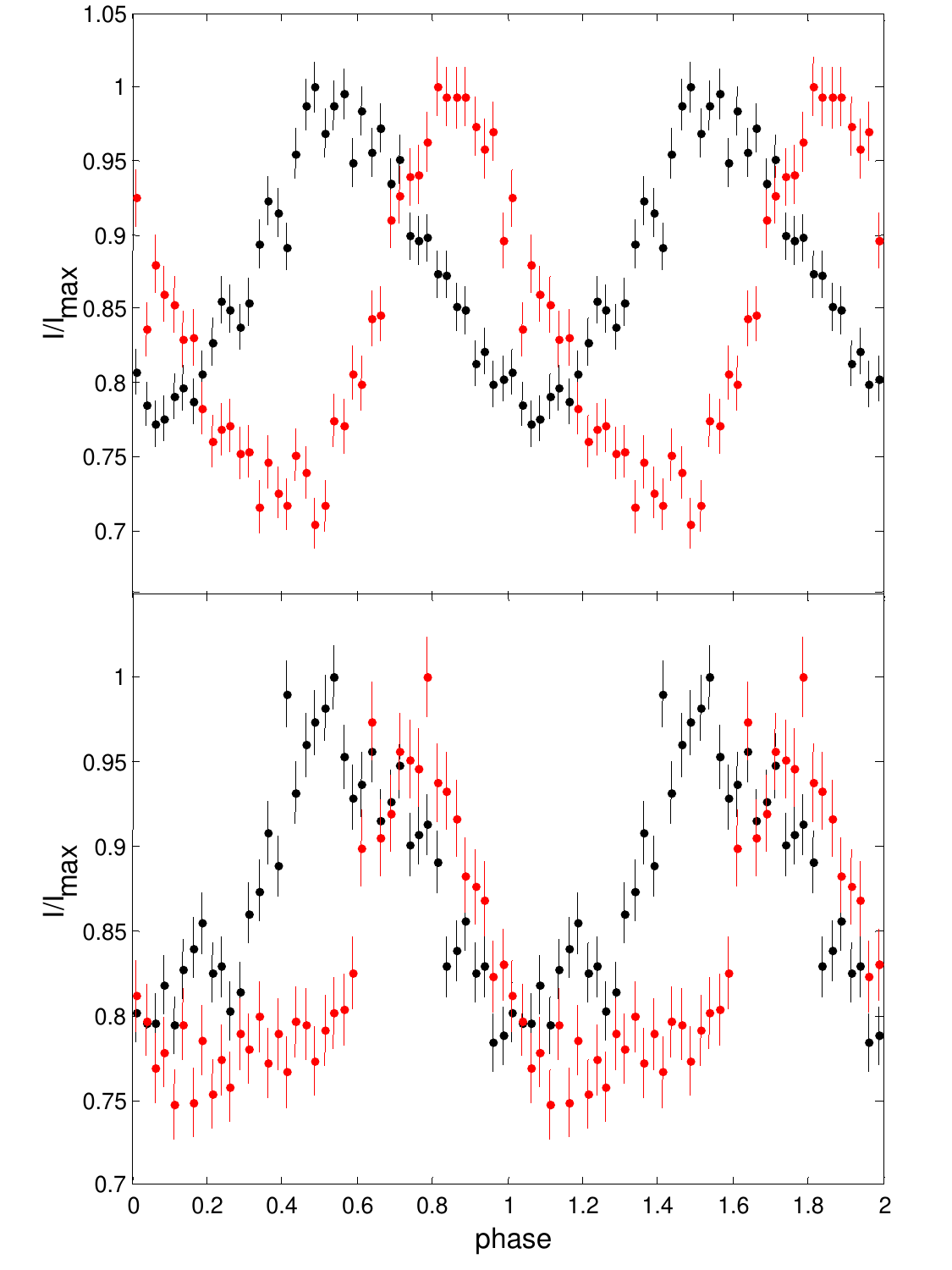}
}
   \caption{Pulse profiles of satellite orbit 622 (black) and 711 (red) in soft (upper panel) and hard (lower panel) bands as an example for the large phase shift applying the timing solution from vK07 with a glitch. The phase residuals are much larger than for the other timing solutions (discussed in the text) and the pulse profile looks more skewed. Error bars denote Poissonian errors. On the vertical axis are the counts are normalized to their maximum value.}
              \label{ReNo0622_0711_K07glitch}
\end{figure}

%Therefore we also fitted a sawtooth to the pulse profile and obtained very similar numbers as in Fig. \ref{phaseres-redsoftblackhard_KuK05} for the phase residuals in the two bands (see Fig. \ref{phaseres_sawtooth_KK05}. The phase shift between hard and soft photons seems even larger. Both model fits have an equal validity regarding to $\chi^{2}$.}
The timing solution of vK07 with a glitch at MJD=52817 days produces phase residuals three times larger than those seen in Fig. \ref{phaseres-redsoftblackhard_KuK05} with a large scatter and will not be considered any further. The pulse profiles for EPIC$-$pn satellite orbit 622 and 711 in both energy bands in Fig. \ref{ReNo0622_0711_K07glitch} are an example of the large phase shifts from one observation to the next by applying this solution.

\begin{table*}%[!tbp]
   \centering
     \caption{\xmm\ EPIC observations of \rxj.}
        \label{xmm-obs}

%\multicolumn{3}{Setup$^{1}$}

\begin{tabular}{|rlll|lll|l|}

\hline
\hline
Orbit	& Observation & Date & MJD & \multicolumn{3}{c}{EPIC instrument setup$^{1}$} & Duration [s]\\
        &             &     & [days] & pn & MOS1 & MOS2                                &(EPIC-pn)\\

\hline
  78 & 0124100101 & 2000 May 13			& 51677 & FF thin & FF thin & SW thin & 62498\\
 175 & 0132520301 & 2000 Nov. 21-22 & 51869 & FF medium & SW thin & - & 26118\\
 533 & 0156960201 & 2002 Nov. 6-7 	& 52584 & FF thin & FF thin & FF thin & 28373\\
 534 & 0156960401 & 2002 Nov. 8-9 	& 52586 & FF thin & FF thin & FF thin & 30173\\
 622 & 0158360201 & 2003 May 2-3 		& 52761 & SW thick & FF thin & FF thin & 72796\\
 711 & 0161960201 & 2003 Oct. 27 		& 52939 & SW medium & FF thin & FF thin & 43013\\
 815 & 0164560501 & 2004 May 22-23 	& 53147 & FF thin & FF thin & FF thin & 43951\\
 986 & 0300520201 & 2005 April 28 	& 53488 & FF thin & SW thin & SW thin & 51435\\
1060 & 0300520301 & 2005 Sep. 23 		& 53635 & FF thin & SW thin & SW thin & 51135\\	
1086 & 0311590101 & 2005 Nov. 12-13 & 53686 & FF thin & SW thin & SW thin & 37835\\
1181 & 0400140301 & 2006 May 22 		& 53877 & FF thin & SW thin & SW thin & 20035\\     
1265 & 0400140401 & 2006 Nov. 05 		& 54044 & FF thin & SW thin & SW thin & 20035\\
1356 & 0502710201 & 2007 May 05 		& 54225 & FF thin & FF thin & FF thin & 20035\\
1454 & 0502710301 & 2007 Nov. 17 		& 54421 & FF thin & FF thin & FF thin & 23059\\

\hline
\multicolumn{8}{l}{}\\
\multicolumn{8}{l}{$^{(1)}$Read - out mode and filter; FF; Full Frame; SW; Small Window.}
\end{tabular}
\end{table*}

\section{Modeling the long term evolution of the timing properties}

From the first \xmm\ observations of \rxj\ Cropper et al. (\cite{cropper1}) reported a phase shift between X-ray 
intensity and hardness ratio (HR, hard and soft band count ratio). Furthermore, de Vries et al. (\cite{Vries04})
found that the phase shift 
changed sign between the year 2000 (the HR leads the intensity profile) and 2003 (the HR lags behind the intensity). As can be seen in Fig. \ref{phaseres-redsoftblackhard_KuK05} this occurred around MJD=52800 days. Since then, the hard band emission lags behind the soft band emission and the shift seems to decrease again from a maximum of $\sim$ 0.1. The phase lag dependence on the energy bands clearly shows that using data from instruments with different energy-dependent sensitivity may introduce systematic errors in the timing analysis. The EPIC-pn instrument has a high 
low-energy response relative to the hard band, therefore Chandra ACIS and EPIC-MOS are expected to be closer to the EPIC-pn 
results restricted to the hard band. 

\begin{figure}[!tbp]
   \centering
   \subfloat{
   \includegraphics[viewport=85 254 490 572, width=0.48\textwidth]{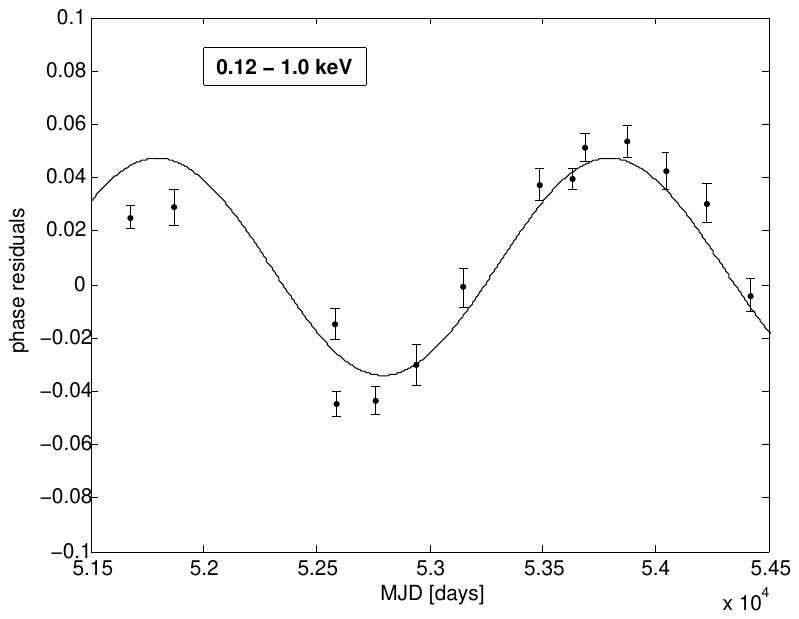}
   }\\
   \subfloat{
   \includegraphics[viewport=85 254 490 572, width=0.48\textwidth]{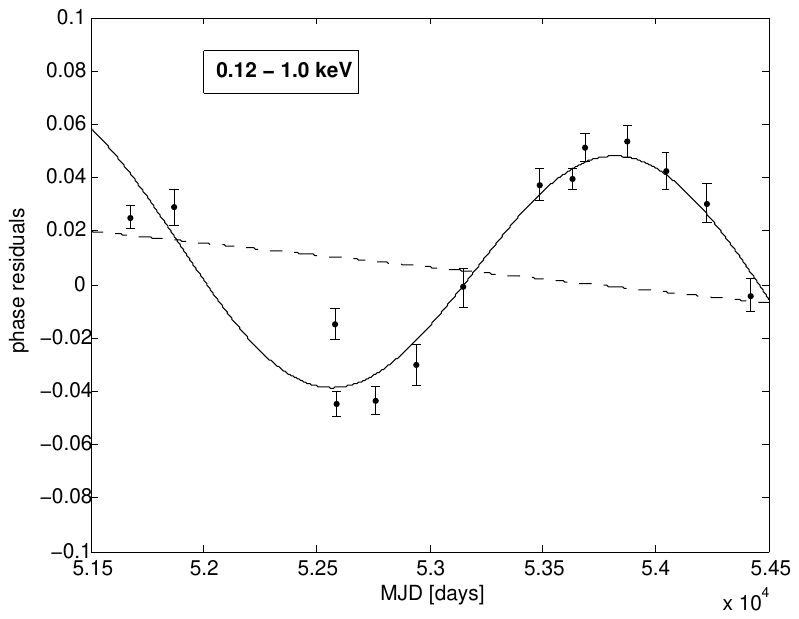}
   }\\
   \subfloat{
   \includegraphics[viewport=85 254 490 572, width=0.48\textwidth]{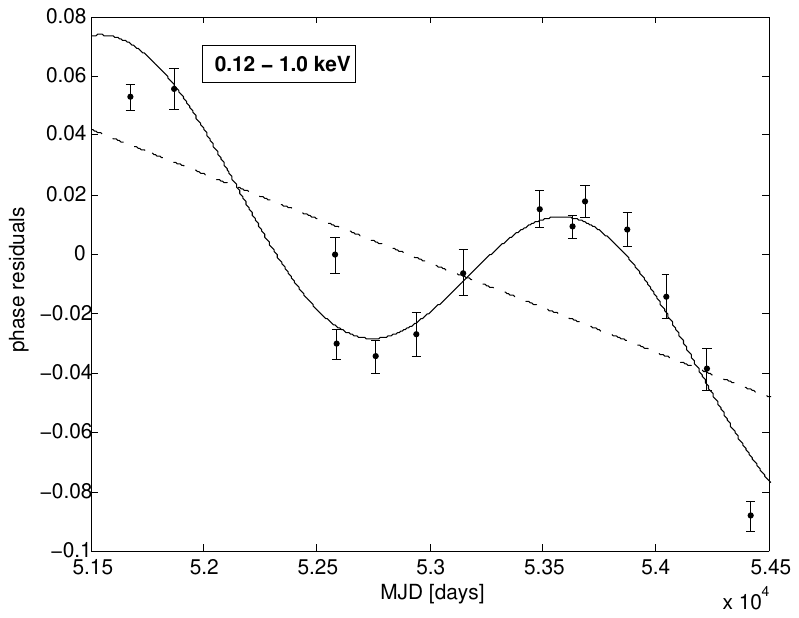}
   }
   
\caption{Phase residuals of \rxj\ derived from the EPIC-pn data using the timing solution from Kaplan \& van Kerkwijk (\cite{KuK05}) 
         fitted with a pure sinusoidal model (top) and a sinusoidal model including a linear term (middle). The fit 
	 with the latter model for the timing solution of van Kerkwijk et al. (\cite{KuK07}) is shown in the bottom panel. 
	 The periods are (5.48 $\pm$ 0.33), (7.20 $\pm$ 1.50) and (5.59 $\pm$ 0.57) yr, respectively, while the values for $\chi^{2}$/d.o.f are
	 59.74/10 (top), 43.05/9 (middle) and 57.25/9 (bottom panel).}
              \label{residuals_120_1000fit}%
\end{figure}

As in previous work, we start modeling the phase residuals using a sine wave function (Fig.~\ref{residuals_120_1000fit}, top).
To avoid systematic effects we use only the EPIC-pn data (120-1000 eV) for our timing analysis.
The different period derivatives, $\dot{P}$, given in KvK05 and vK07 depend on which and how many data points and from which instruments they were used. We already mentioned the different energy response of the different instruments. $\dot{P}$ depends on how much of the cycle, shown in Fig.~\ref{phaseresall}, is covered with observations and can be determined more accurately only in the future, when at least one full cycle of the variation of the phase residuals is observed, assuming a 10-12 yr period as discussed below. Therefore, we introduced an additional linear term y=mt+n, as can be seen in Fig.~\ref{residuals_120_1000fit} for the timing solution of KvK05 (middle) and that of vK07 with constant spin-down without glitches (bottom). 

The two maxima in the phase residuals (Fig.~\ref{residuals_120_1000fit}) do not necessarily have equal heights. By adding 
the linear term, we automatically take into account the different values of $\dot{P}$ and a possible different 
amplitude of the two maxima covered by the observations. For the three cases described above, we derive periods 
of (5.58 $\pm$ 0.33) yr, (7.20 $\pm$ 1.50) yr and (5.59 $\pm$ 0.57) yr, respectively (here and below, errors on the periods correspond to 1$\sigma$). The values are consistent within 
their errors and in particular the first and last value are nearly identical. This is expected because a possible period should not significantly depend on the actual timing solution (as long as this is correct to first order). It should be noted, however, that the fits with a sine wave are formally not acceptable, given the derived $\chi^{2}$ values. The unacceptable fit can be partly explained by the two data points from MJD=52584 and 52586~ days. It suggests that the sine wave is not a good representation of the data. We excluded the data point from MJD=52584~ days and derived periods of (6.54 $\pm$ 0.14), (6.55 $\pm$ 0.46) and (5.73 $\pm$ 0.43) yr, respectively, for the three cases, while the values for $\chi^{2}$/d.o.f are 17.16/9, 16.90/8 and 35.12/8, respectively. Only the fit with the data shown in Fig. ~\ref{residuals_120_1000fit} (top), excluding the data point at MJD=52584~ days, is formally acceptable. The reason for the outlier at this epoch is unclear.

In the first observation the soft photons lag behind the hard photons, while in the observations after MJD=52800 days the opposite 
is true (see Fig. \ref{phaseres-redsoftblackhard_KuK05}). The phase lag in the first observation is supported by Fig. 2 of de~Vries~et~al. (\cite{Vries04}), which is independent of a model fit to the pulse profile. The energy dependent heights of the two phase residual 
maxima (Fig. \ref{phaseres-redsoftblackhard_KuK05}) show that the actual period of the modulation could be twice the sine 
period. The energy dependence suggests a model of emission originating from two hot spots with different temperature. 
In the case of a precession model, for one half of the precession period the soft emission (cooler pole) would
precede the hard emission and for the other half it would be 
the other way around. The cooler cap would be mostly visible for half the precession cycle, while in the other half the hotter cap would dominate emission. If the emission arises in different poles this suggests that the profile should be that 
of an abs(sine), similar to that used by vK07 and van Kerkwijk \& Kaplan (\cite{vKK07}) but with a different interpretation. They found a period $\approx$4.3 yr as a solution equivalent to the glitch model. Unfortunately van Kerkwijk \& Kaplan (\cite{vKK07}) do not provide errors, and we cannot check if our value of the period $\approx$5.5 yr is consistent with theirs. The difference of the values may be explained by the four new EPIC-pn observations we add.

\begin{figure}[!tbp]
   \centering
   \subfloat{
   \includegraphics[viewport=85 254 490 572, width=0.48\textwidth]{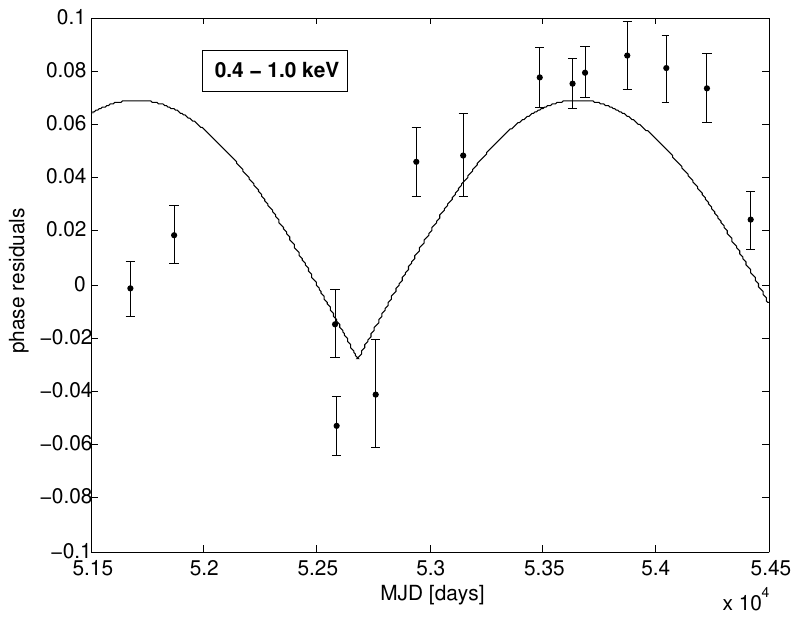}
   }\\
   \subfloat{
   \includegraphics[viewport=85 254 490 572, width=0.48\textwidth]{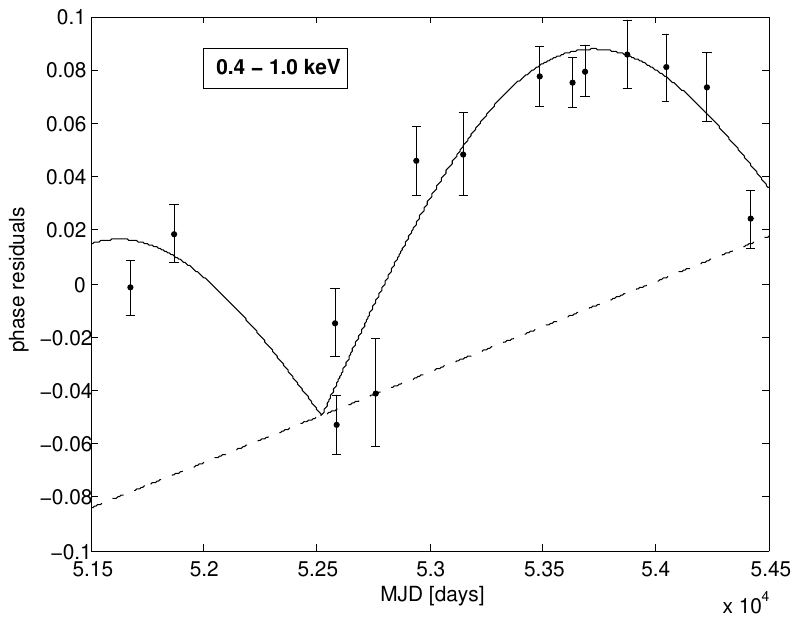}
   }\\
   \subfloat{
   \includegraphics[viewport=85 254 490 572, width=0.48\textwidth]{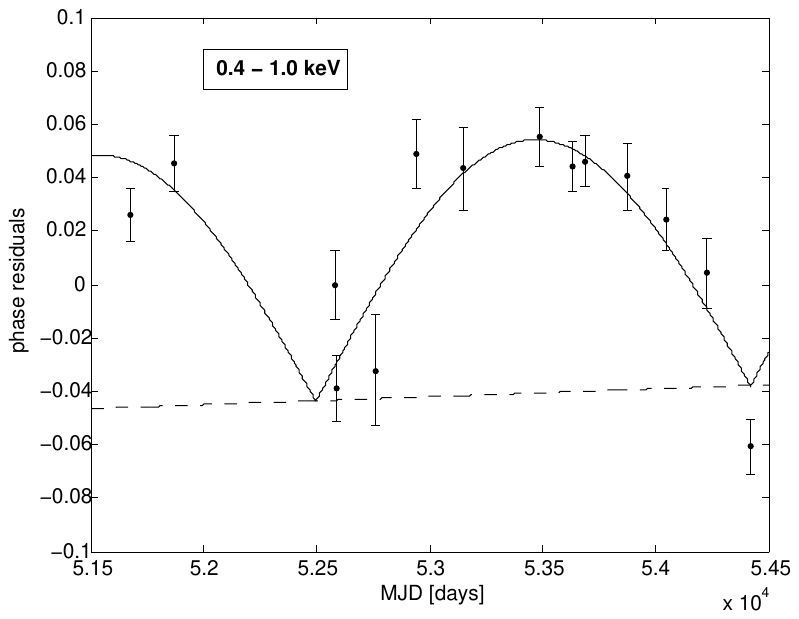}
   }
   
\caption{Phase residuals of \rxj\ derived from the EPIC-pn 0.4-1.0 keV data.
         The error of the phase residual from revolution 622 (around MJD=52800 days) was increased by a 
	 factor of two (see text). 
         Shown are the best fit models based on different abs(sine) functions and different timing solutions
	 as summarized in Table~\ref{KapSou}. 
	 The derived periods are (10.711 $\pm$ 0.058) yr (model B1, top), (11.504 $\pm$ 0.072) yr (B2, middle) 
	 and (10.546 $\pm$ 0.069) yr (B3, bottom).}
    \label{residuals_120_1000fit_abs}
\end{figure}

We interpret the turn-around in the phase residual evolution (near MJD=52800 days) not as a glitch, but as "switching" of the emission 
to the predominantly visible pole. As in the fit with the normal sine function, we again add the linear term to the 
abs(sine). To reduce systematic effects due to the use of different filters we use the EPIC-pn hard band (400-1000 eV)
data. We use the hard band in order to be able to include data from other instruments, although statistics in the soft band would be preferable. For the fit we further increase the size of the error of the phase residual in revolution 622 by a factor of two (model B1-B4). 
This observation was performed with the thick filter, which has a significantly lower transmission in the soft energy band compared to 
the medium and thin filter. This causes a systematic shift of this phase residual to larger values and would introduce 
systematic errors in the fit results. Excluding this observation completely (model A1-A4), we obtained a period which is 
2 - 4.5 years longer (see Table~\ref{KapSou}). To get an estimate of the systematic error of the 
period, which is likely larger than the statistical error derived from the fit, we also applied a 
doubled error for all observations which were not performed with the thin filter, i.e. revolutions 
175, 622 and 711 (model C1-C3). The results of the fitting procedures are listed 
in Table~\ref{KapSou} and plotted in Fig. \ref{residuals_120_1000fit_abs}. Except for model A, the advantage of introducing an additional slope is clearly visible in the $\chi^{2}$/d.o.f. as compared to the poor fits without a slope.

Applying a sine instead of the abs(sine) models B1, B2 and B3, we derive $\chi^{2}$/d.o.f.=3.31, 3.67 and 4.85, compared to 12.25 (top), 2.43 (middle) and 4.99 (bottom) in Fig.~\ref{residuals_120_1000fit_abs}. The fit of model B1 is much better with a sine instead of an abs(sine) but a better approximation would be an abs(sine) with an additional slope, which gives the best $\chi^{2}$=2.43 (B2) in this case for the timing solution of KvK05. The two $\chi^{2}$ values for the timing solution of vK07 are not significantly different.
 
\begin{figure}
  \centering
   \resizebox{\hsize}{!}
{
   \includegraphics[viewport=85 254 490 572, width=0.48\textwidth]{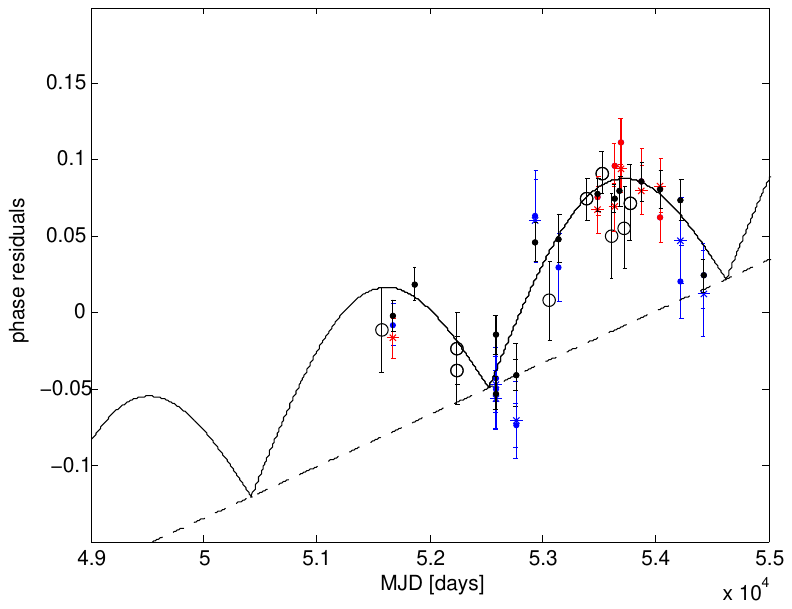}
}
   \caption{The phase residuals (only hard band) shown in Fig. \ref{allphaseres_chandra_all} together with model B2 derived from the EPIC-pn (hard) data shown in Fig. \ref{residuals_120_1000fit_abs} (middle). Phase residuals from ROSAT and Chandra with errors larger than 0.03 (phase residuals) are excluded.}
              \label{phaseresallfitabssineposgerade}%
    \end{figure}

For a consistency check we show in Fig. \ref{phaseresallfitabssineposgerade} the abs(sine) model fit shown in 
Fig. \ref{residuals_120_1000fit_abs} (middle) together with the phase residuals obtained from EPIC-pn, EPIC-MOS and 
ACIS (all hard band) and HRC observations. We excluded in this Figure Chandra data points with errors 
$\geq$ 0.03.

\begin{table*}
   \centering
     \caption{Periods derived from the fits of the phase residuals with a model consisting of abs(sine)+mt+n for
              different timing solutions. The plots for models B1-3 are shown in Fig. \ref{residuals_120_1000fit_abs}. All errors correspond to 1$\sigma$.}
        \label{KapSou}
\begin{tabular}{|lllllllll|}
\hline
\hline
Model & Timing    & period & m & amplitude & offset (n) & $\chi^{2}$ & d.o.f & $\chi^{2}$/d.o.f \\
      & solution  & [yr]   & [phase res./s] & [phase res.] & [phase res.] & & \\
\hline

\multicolumn {9} {|c|} {revolution 622 excluded}\\
\hline
A1 & KvK05 & 15.21  $\pm$ 0.65 & 0 & 0.163 $\pm$ 0.068 & -0.079 $\pm$ 0.017 & 24.51 & 9	&	2.72\\
A2 & KvK05 & 15.210 $\pm$ 0.011 & 7.45 $\cdot 10^{-11}$ & 0.15 $\pm$	0.36 & -0.41 $\pm$ 3.40 & 22.75 &	8	&	2.84\\
A3 & vK07 & 12.488 $\pm$ 0.055 & -1.10 $\cdot 10^{-11}$ & 0.125 $\pm$	0.034 & -0.01	$\pm$	0.41 & 64.49 & 8 &	8.06\\
A4 & vK07 & 10.693 $\pm$ 0.059 & 0 & 0.065 $\pm$	0.028 & -0.015 $\pm$	0.015 & 61.53 & 9 & 6.84\\
\hline

\multicolumn {9} {|c|} {doubled error of revolution 622}\\
\hline
B1 & KvK05 & 10.711 $\pm$ 0.058 & 0 & 0.097 $\pm$ 0.050 & -0.028 $\pm$	0.035 & 122.45 & 10	& 12.25\\
B2 & KvK05 & 11.504 $\pm$ 0.072 & 39.21 $\cdot 10^{-11}$ & 0.099	$\pm$	0.029 & -1.83	$\pm$	0.40 & 21.87 & 9 & 2.43\\
B3 & vK07 & 10.546 $\pm$ 0.069 & 3.44 $\cdot 10^{-11}$ & 0.095 $\pm$ 0.034 & -0.20 $\pm$ 0.42 & 44.89 & 9 & 4.99\\
B4 & vK07 & 9.656 $\pm$ 0.023 & 0 & 0.057 $\pm$ 0.026 & -0.012 $\pm$ 0.017 & 51.38 & 10 & 5.14\\ 
\hline
\multicolumn {9} {|c|} {doubled error of revolution 175, 622 and 711}\\
\hline

C1 & KvK05 & 9.07   $\pm$ 0.14  & 0 & 0.058	$\pm$	0.068 & 0.003 $\pm$	0.042 & 138.33 & 10 &	13.83\\
C2 & KvK05 & 9.376  $\pm$ 0.037 & 45.17 $\cdot 10^{-11}$ & 0.087 $\pm$	0.035 & -2.09	$\pm$	0.48 & 18.45 & 9 & 2.05\\
C3 & vK07 & 9.009  $\pm$ 0.037 & 2.48 $\cdot 10^{-11}$ & 0.090	$\pm$	0.040 & -0.15 $\pm$	0.53 & 32.61 & 9 & 3.62\\
C4 & vK07 & 9.017  $\pm$ 0.038 & 0 & 0.076	$\pm$	0.031 & -0.025 $\pm$	0.019 & 45.00 & 10 & 4.50\\
\hline
\end{tabular}
\end{table*}

%\begin{figure}
%  \centering
%   \resizebox{\hsize}{!}
%{
%   \includegraphics[viewport=85 254 490 572, width=0.48\textwidth]{phaseresallfit.pdf}
%}   %%%\includegraphics{empty.eps}
%   \caption{The phase residuals shown in Fig. \ref{phaseresall} fitted with a sine wave. ROSAT data is excluded, the period is (5.83 $\pm$ 0.25) yr}
%              \label{phaseresallfit}%
%    \end{figure}

\section{The long term spectral evolution}
To avoid systematic errors due to cross-calibration uncertainties between the different instruments we 
restrict our spectral analysis to the EPIC-pn data. Moreover we use only the eleven EPIC-pn 
observations of \rxj, which were done in full frame mode with the thin filter. We extracted 
the pulse-phase averaged spectra from circular source and background regions with 1\arcmin\ in diameter, 
selecting single-pixel events (PATTERN = 0). The spectra were binned to obtain at least 100 photons in each bin. 
The fits were performed with XSPEC12 and restricted to energies between 0.16 keV and 1.5 keV.
We used the blackbody model plus an additive Gaussian line, which represents the absorption 
feature (negative normalisation). 
The energy resolution of the instrument is not sufficient to determine the shape of the absorption 
feature. A Gaussian profile applied multiplicatively yields equally good fits. The feature might also
consist of several unresolved lines or have the shape of an absorption edge. Here and below, the errors on the spectral parameters correspond to the 2.7$\sigma$ confidence level. 

We linked all model parameters for the spectra of revolution 533 and 
534, because the two observations were only two days apart. The line energy, the line width $\sigma$
and the hydrogen column density $N_{H}$ were linked for all spectra, i.e. we assume that they do not
vary with time. All spectra were fitted simultaneously. 
The best fit parameters are (301 $\pm$ 3) eV for the line energy, $\sigma$ = (77 $\pm$ 2) eV
and $N_{H}$ = (1.04 $\pm$ 0.02)$\cdot10^{20}$ cm$^{-2}$. For all spectra together we obtain $\chi^{2}$/d.o.f$=$1.29 with 1767 degrees of freedom. The blackbody temperature kT, the blackbody normalisation (used to calculate the size of the emitting area, assuming a distance of 300 pc) and the line equivalent width EW were free parameters and their best fit values are listed in Table~\ref{pn-spectral}.
The time evolution of the parameters is drawn in Fig.~\ref{ktewr}. We also included kT and EW of the three EPIC-pn observations done with different observing mode and different instrument setup in this figure, which were not used for this fit. Because of the cross-calibration uncertainties mentioned above, the blackbody radii obtained from these observations are not reliable and thus not included in Fig.~\ref{ktewr}, bottom. 

As can be seen in Fig.~\ref{ktewr}, the sine wave form is not a good approximation for the new data. We only 
can conclude that, if the variation in the spectral parameters is cyclic, the period is larger than 
the time span of the EPIC-pn observations. Currently it is not possible to predict whether it is 
periodic or not and if it is periodic, what the shape of the variation might be.

Following H06 we computed pulse phases for each detected event using the timing solution 
of KvK05. We divided one pulse period into five phase intervals and extracted spectra from each phase interval
for each observation. As in the analysis of the phase-averaged spectra, we performed a joint fit to
the 55 spectra with the same parameters linked ($\chi^{2}$/d.o.f$=$1.16 with 6452 degrees of freedom).
The change of kT and EW during the pulse period for each of the observations is shown
in Fig.~\ref{EW_vs_kT_5phaseKuK05}. As seen before, the parameters evolve counter-clockwise and the long term
trend of the parameters, as seen in Fig.~\ref{ktewr} continues at all pulse phases.
%In Fig.~\ref{fluxbands} we show that the hardness of the spectrum is changing compared to Fig.~\ref{ktewr} 
%top correlating to the temperature. For the first two EPIC-pn observations the soft photons dominated the 
%flux, while the hardness of the flux changed around MJD=52800 days. This change occurred at the same time 
%the phase residuals had their minimum (see Fig.~\ref{phaseresallfitabssineposgerade}). Now the flux seems 
%to become softer again and the phase residuals aspire to a new minimum. This matches the assumption of 
%two hot spots.

%\section{Short term variation}

\section{Discussion}

We present new \xmm\ observations of the isolated neutron star \rxj\ performed during our half-year 
monitoring program in 2006 and 2007. Combining the new data with the previous \xmm\ data 
presented by H06 and with Chandra data (KvK05, vK07), we followed the spectral and timing behaviour 
of the source.
As shown by KvK05 and vK07, a phase-coherent timing analysis assuming a constant spin-down of the pulsar
yields relatively large phase residuals. Different models have been fitted to these phase residuals
invoking a more sudden change in properties (vK07) on one hand or a periodic behaviour 
(H06 and vK07) on the other hand. Physically, these effects could be interpreted, 
e.g., by a ``glitch'' or by precession of the neutron star, respectively.

\begin{table}
   \centering
     \caption{All EPIC-pn observations in full frame mode with the thin filter of \rxj\ with the derived variations of the pulse averaged spectra. Observations done in different observing modes, which were not used for the fit are in italic. The radius obtained from observation 711 is not used at all (see explanation at the end of section 5). All errors correspond to 2.7 $\sigma$ confidence level.}
        \label{pn-spectral}
\begin{tabular}{ccccc}
\hline
\hline
Orbit	 &MJD     &	kT	     &EW	&	radius\\
	 &	  &	[eV]	     &	[eV]	&	[km]\\
 \hline
78	 &51677   & 86.5 $\pm$ 0.4 &  -4.0 $\pm$ 4.0 & 4.79 $\pm$ 0.09\\ 
\it{175} &\it{51870}   & \it{86.0 $\pm$ 0.7} &  \it{+5.6 $\pm$ 7.0} & \it{4.90 $\pm$ 0.14}\\
533/534  &52585/7 & 88.4 $\pm$ 0.4 & -16.8 $\pm$ 3.3 & 4.70 $\pm$ 0.07\\
\it{711} &\it{52940}   & \it{92.0 $\pm$ 0.6} &  \it{-64.7 $\pm$ 5.4} & \it{$-$}\\
815	 &53148   & 94.6 $\pm$ 0.5 & -58.6 $\pm$ 4.2 & 4.29 $\pm$ 0.08\\
986	 &53489   & 94.3 $\pm$ 0.4 & -57.8 $\pm$ 3.8 & 4.35 $\pm$ 0.07\\
1060	 &53636   & 93.7 $\pm$ 0.5 & -55.0 $\pm$ 3.8 & 4.37 $\pm$ 0.08\\
1086	 &53687   & 93.1 $\pm$ 0.5 & -55.9 $\pm$ 3.8 & 4.50 $\pm$ 0.08\\
1181	 &53877   & 93.1 $\pm$ 0.6 & -51.2 $\pm$ 4.8 & 4.44 $\pm$ 0.10\\
1265	 &54045   & 92.7 $\pm$ 0.6 & -53.6 $\pm$ 4.7 & 4.53 $\pm$ 0.11\\
1356	 &54226   & 92.4 $\pm$ 0.6 & -44.1 $\pm$ 4.9 & 4.45 $\pm$ 0.11\\
1454	 &54421   & 91.8 $\pm$ 0.6 & -44.6 $\pm$ 4.6 & 4.50 $\pm$ 0.10\\
\hline
\end{tabular}
\end{table}

 \begin{figure}[!tbp]
   \centering
   \subfloat{
   \includegraphics[viewport=85 254 490 572, width=0.48\textwidth]{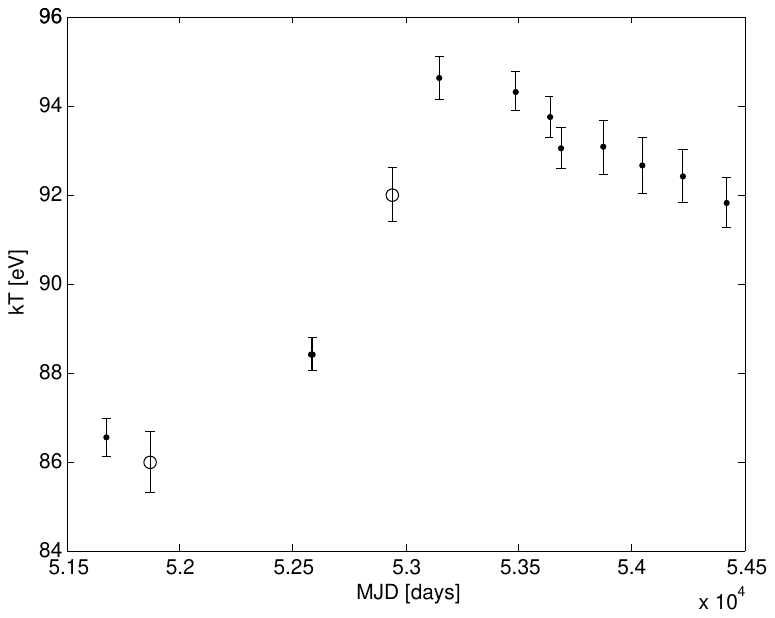}
   }\\
   \subfloat{
  \includegraphics[viewport=85 254 490 572, width=0.48\textwidth]{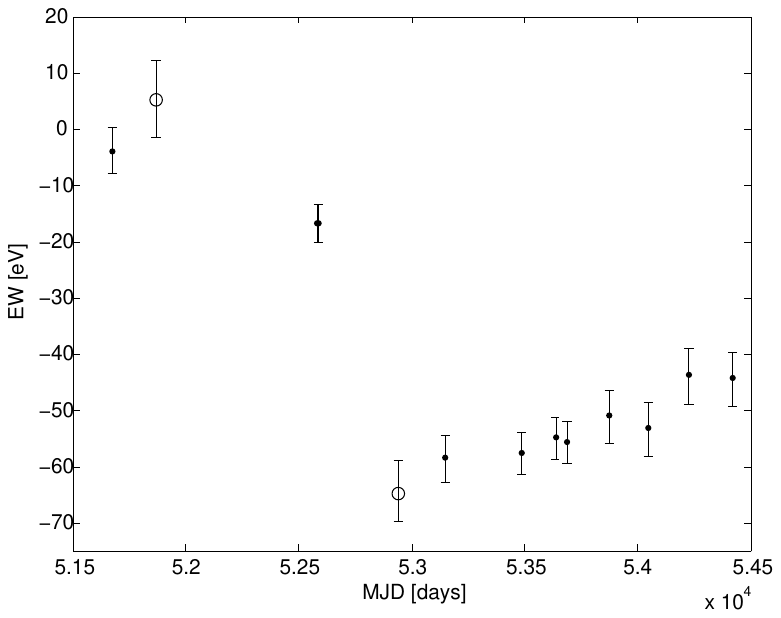}
   }\\
   \subfloat{
   \includegraphics[viewport=85 254 490 572, width=0.48\textwidth]{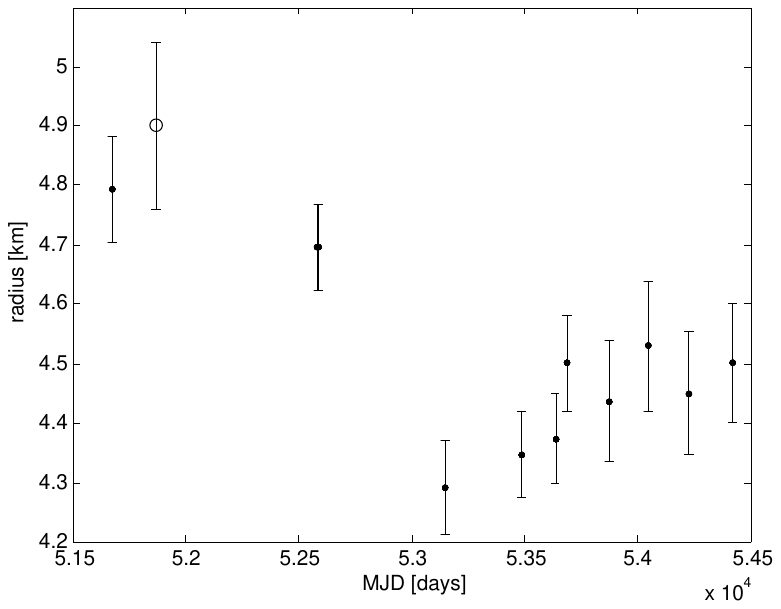}
   }
   
\caption{Long term variations of the pulse phase averaged spectra obtained from the eleven EPIC-pn observations in full frame mode with a thin filter (dots). Open circles show data obtained from the other EPIC-pn observations (see Table \ref{xmm-obs}). Error bars denote 2.7$\sigma$ confidence level.}
              \label{ktewr}%
\end{figure}

\begin{figure}
  \centering
   \resizebox{\hsize}{!}
{
   \includegraphics[viewport=85 254 490 572, width=0.48\textwidth]{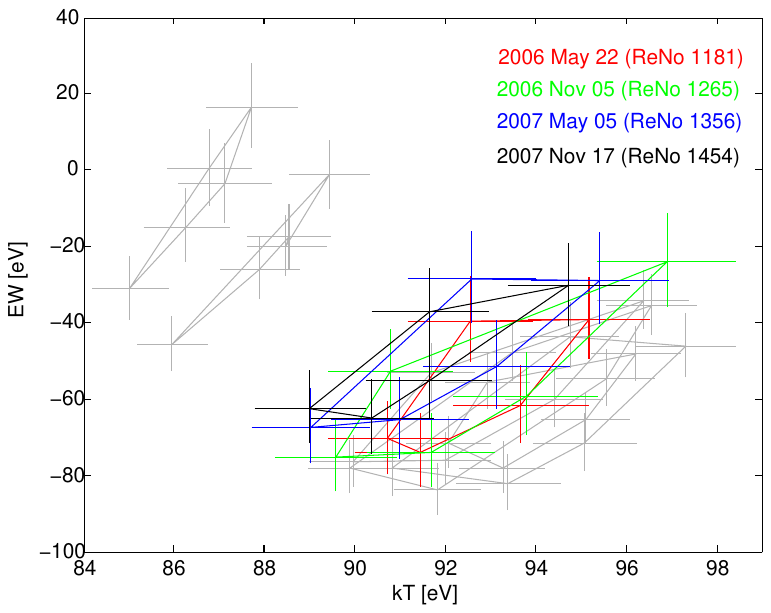}
}   
   \caption{Phase resolved variations of equivalent width versus temperature. Observations already used for Fig. 6 in H06 are marked in grey. The four new EPIC-pn data are coloured. The error bars denote 2.7$\sigma$ confidence level.}
             \label{EW_vs_kT_5phaseKuK05}%
\end{figure}

The analysis of all available data from the different instruments on \xmm\ and Chandra yields 
systematic shifts between the phase residuals (Fig.~\ref{phaseresall}). As already noted by 
Haberl (\cite{Hab07}) this is most likely caused by intrinsicaly energy dependent phase residuals and this must be accounted for when using different instruments of different energy response. This is confirmed by our timing analysis using 
events from the EPIC-pn data in the soft (120 - 400 eV) and hard (400 - 1000 eV) energy 
band separately, which yield systematic phase shifts (Fig.~\ref{phaseres-redsoftblackhard_KuK05}).
Moreover, the phase shifts between soft and hard photons vary with time.
In order to still combine the results from the different instruments, we restricted our analysis
to the hard energy band for the three EPIC instruments and for ACIS. The negligible spectral 
capabilities of the HRC prevented an energy selection and we excluded the ROSAT data completely
due to the large errors and alias periods caused by the interrupted observations.
This brought the phase residuals from the various instruments in much closer agreement
(Fig.~\ref{allphaseres_chandra_all}). 

Because of the systematic shifts of the pulse phases on time scales of years, which become visible 
as phase residuals with respect to the model with constant spin-down of the neutron star, the 
solutions for $\dot{P}$ strongly depend on the time interval used for the analysis.
Only when the observations eventually cover a period significantly longer than the time scale of the 
phase variations - which is currently not the case - a reliable value for $\dot{P}$ can be derived. This is true if the phase variations are periodic, but also for a sudden event with a long ($\geq$10 yrs) relaxation time.
To overcome this problem, we included a linear term into our long term modeling of the phase
residuals and investigate two solutions for $\dot{P}$ as inferred by KvK05 (all data solution) 
and vK07 (all data solutions without glitches). However, we emphasize that both solutions
should be regarded as first approximations and that any future timing analysis combining data from 
different instruments needs to be restricted to narrower energy bands.

Using an abs(sine) function to model the variations of the phase residuals, we obtain periods 
of 9-15 years with the most reliable values around 9.3-11.5 years for the two fits with the lowest $\chi^2$ (Table~\ref{KapSou}). 
The observations still cover less than two ``humps'' of the abs(sine) function and more data are needed
to ultimately verify that the timing behaviour is indeed cyclic and to determine the period more 
precisely. 
The energy dependence of the phase shifts (Fig.~\ref{phaseres-redsoftblackhard_KuK05}) clearly 
demonstrates that the long term period covers two humps of the abs(sine). This was not taken into 
account previously and is at least partly the reason for the shorter periods determined in the past 
(H06, vK07). 

For the spectral analysis of \rxj\ it is even more important to avoid systematic differences due to 
cross-calibration uncertainties between the various instruments. Therefore, we restricted our 
spectral analysis to the EPIC-pn data taken in full frame CCD readout mode and with a thin 
optical blocking filter. As in H06 we use a simple absorbed blackbody
model including an absorption feature which we assume to be of Gaussian shape.
The new \xmm\ observations show that the inferred spectral parameters continue to change, but
do not follow the sinusoidal evolution suggested by the previous subset of data used in H06.
Temperature, kT, the apparent radius of the emitting area and the equivalent width, EW, of the 
absorption line changed more slowly over the years 2006 and 2007 than an extrapolation of the 
sine model would predict. The spectral data do not allow us to draw any conclusions about periodicity. 

The phase lags between soft and hard emission (Fig.~\ref{phaseres-redsoftblackhard_KuK05}) suggest
the existence of at least two hot regions on the neutron star surface with 
somewhat different emission 
characteristics. These hot spots might be associated with the polar caps of the neutron star
in a magnetic field. Two hot spots have also been suggested to explain the
rotational variations seen in the light curves of RBS\,1223 - another M7 star - in 
different energy bands (\cite{2005A&A...441..597S}). To explain the different energy 
dependence of the two peaks are seen in 
the pulse profiles of RBS\,1223, these authors suggest two hot spots with different size and 
temperature which are not anti-podal. A similar geometry
can in principle also account for the spectral variations seen in \rxj\ during its spin period.
However, to explain the change in the phase lags over time seen in \rxj, requires a slow
apparent movement of the spots relative to each other as seen from the distant observer.

One possible explanation would be free precession of the neutron star as suggested by 
de Vries et al. (\cite{Vries04}) and put forward by H06. In this case the viewing geometry 
changes and no physical movement of the hot spots on the stellar surface is required. This 
model is discussed in detail in H06, but as already discussed above, can only be verified 
when the periodic behaviour is confirmed. The simple, free precession model discussed in H06 was able to account (at least qualitatively) for the long-term evolution of the spectral parameters over the period examined (May 2000-November 2005) and also by invoking two non-antipodal caps for the peculiar intensity-hardness anti correlation with spin phase. Although this was not discussed in H06, the model predicts a cyclic variation in the phase residuals due to the change in the pulse shape (of the phase of the light curve maximun in particular) over the precession period. Since the spin-phase dependence of the hardness ratio changes along the precession cycle (this can be also seen in the lower right panel of Fig. 5 in H06), one expects that the phase residuals and their variation over time are not the same in different energy bands. Indeed, as we checked, the free precession model outlined in H06 is able to reproduce (again, qualitatively) the properties of the observed residuals (see Fig. \ref{phaseres-redsoftblackhard_KuK05}). The highly non-sinusoidal pattern of the spectral parameters, which clearly emerges from the latest data (see Fig. \ref{ktewr}), however, may be, an issue. In order to explain the fast change around MJD = 52800 days one needs to invoke the coming into view of a rather small ''spot'' with angular size $\sim \Delta t/P_{prec}\approx 10^\circ$ where we take $\Delta t\sim 500$ d as the timescale over which the rapid change occurred and $P_{prec}\sim 10$yr. However, the presence of such a small emitting region produces a pulse shape which is not in agreement with the observed one. It is still under debate whether free precession 
can continiue for more than a few hundred cycles (Link \cite{link}). If this is the case
and \rxj\ is precessing, then the precession needs to be powered by some mechanism. 
Precession caused by an unseen sub-stellar companion can probably be excluded on the basis of the current 
upper limit on its mass obtained by Posselt et al. (\cite{posselt}) from a search for orbiting
objects around \rxj\ (see also the discussion on PSR\,B1828-11 by Liu et al. \cite{liu}).

Alternatively, the regions on the surface responsible for the hard and soft emission
may physically change their locations or their emission properties. A slow rearrangement of the magnetic field may change the local angle between the surface normal and the field, influencing the emission pattern, and the heat transport in the crust. In such a picture the rearrangements 
do not need to be periodic, but a different physical mechanism is required to explain it.
A glitch (vK07) may be one possibility. The occurrence of such a sudden event might be supported
by the relatively fast change observed in the spectral parameters around MJD = 52800 days 
(see Fig.~\ref{ktewr}) and the more gradual evolution afterwards which resembles a relaxation 
process. 
However, the EPIC-pn small window mode data from 2003 (satellite revolution 711)
and in particular the RGS data indicate a more gradual temperature increase before 
MJD = 52800 days (de Vries et al., \cite{Vries04}). The small window mode data (thick and medium filter) was not used here at all and only partly by H06, because of a normalisation discrepancy between small window and full frame mode in the EPIC-pn calibration. This affects the inferred black body radius, but not the temperature.
It should also be noted that the basic characteristic behaviour in the spectral variations was
not altered  around this date. The changes in kT, radius and EW seen on long term
time scales are also seen during the neutron star rotation, before and after the putative 
event (Fig.~\ref{EW_vs_kT_5phaseKuK05} and compare to Fig. 4 in H06) and the total X-ray flux stays nearly constant
within a few per cent (H06). The amplitude of the EW variation during the spin period did not change over the years, only the mean level evolves. In the case of the temperature, both amplitude and mean level evolve on long time scale, but this evolution is visible in the EPIC-pn spectra during all observations, first slowly and then fast with a large change around MJD = 52800 days. Therefore, we regard a sudden event like a glitch as an unlikely cause for the (more gradual) spectral 
variations.

In conclusion, the timing and spectral properties of \rxj\ are both consistent with the 
existence of at least two hot spots with different temperature on the surface of the neutron 
star. Whether the long term changes are periodic requires further monitoring. A simultaneous 
modeling of spectral and timing parameters is in progress and should eventually yield a 
more detailed temperature map of this unique isolated neutron star.

\begin{acknowledgements}
The \xmm\ project is supported by the Bundesministerium f\"ur Wirtschaft und
Technologie/Deutsches Zentrum f\"ur Luft- und Raumfahrt (BMWI/DLR, FKZ 50 OX 0001)
and the Max-Planck Society.
MMH and VH acknowledge support by the Deutsche Forschungsgemeinschaft (DFG) through
SFB/TR 7 ``Gravitationswellenastronomie''.
The work of RT is partially funded by INAF-ASI through grant AAE TH-058 and SZ acknowledges the STFC for support through an Advanced
Fellowship.
\end{acknowledgements}

%\newpage
{}


\begin{thebibliography}{}
  \bibitem[2007]{bug} Blaschke, D., Grigorian, H. 2007, PrPNP, 59, 139B
  \bibitem[2001]{cropper1} Cropper, M., Zane, S., Ramsay, G., et al. 2001, A\&A, 365, 302 
  \bibitem[2004]{cropper} Cropper, M., Haberl, F., Zane, S., Zavlin, V.E. 2004, MNRAS, 351, 1099
  %\bibitem[2003]{Vriesvar} de Vries, C. P., Vink, J., M\'endez, M., Verbunt, F. 2003, AJ, 126, 1217
  \bibitem[2004]{Vries04} de Vries, C. P., Vink, J., M\'endez, M., Verbunt, F. 2004, A\&A, 415, L31
  \bibitem[2005]{Hab05} Haberl, F. 2005, MPE Report, 288, 39
  \bibitem[Haberl et~al. 1997]{1997A&A...326..662H}
          {Haberl}, F., {Motch}, C., {Buckley}, D.~A.~H.,et al. 1997, A\&A, 326, 662
  \bibitem[2006]{Hab06} Haberl, F., Turolla, R., De Vries, C. P., et al. 2006, A\&A, 451, L17 (H06)
  \bibitem[2007]{Hab07} Haberl, F. 2007, Ap\&SS, 308, 181
  \bibitem[2001]{jansen} Jansen, F., Lumb, D., Altieri, B., et al. 2001, A\&A, 365, L1
  \bibitem[2002]{KuK02} Kaplan, D. L., Kulkarni, S. R., van Kerkwijk, M. H., Marshall, H. L. 2002, ApJ, 570, L79
  \bibitem[2005]{KuK05} Kaplan, D. L., van Kerkwijk, M. H. 2005, ApJ, 628, L45 (KvK05)
  \bibitem[2007]{Kaplandist1856} Kaplan, D. L., van Kerkwijk, M. H., Anderson, J. 2007, ApJ, 660, 1428
  \bibitem[2007]{KuK07} van Kerkwijk, M. H., Kaplan, D. L., Pavlov, G. G., Mori, K. 2007, ApJ, 659, L149 (vK07)
  \bibitem[2006]{link} Link, B. 2006, A\&A, 458, 881
  \bibitem[2006]{liu} Liu, K., Yue, Y. L., Xu, R. X. 2007, MNRAS, 381, L1
  \bibitem[2006]{page1} Page, D., Reddy, S. 2006, ARNPS, 56, 327
  \bibitem[2006]{page2} Page, D., Geppert, U., Weber, F. 2006, NuPhA, 777, 497
  \bibitem[2008]{posselt} Posselt, B., Neuh\"auser, R., Haberl, F. 2008, A\&A submitted
  \bibitem[Schwope et al. 2005]{2005A&A...441..597S}
          {Schwope}, A.~D., {Hambaryan}, V., {Haberl}, F., \& {Motch}, C. 2005, A\&A, 441, 597
  \bibitem[2001]{strueder} Str\"uder, L., Briel., U., Dennerl, K., et al. 2001, A\&A, 365, L18
  \bibitem[2004]{trümper} Tr\"umper, J. E., Burwitz, V., Haberl, F., Zavlin, V. E. 2004, NuPhS, 132, 560
  \bibitem[2001]{turner} Turner, M. J. L., Abbey, A., Arnaud, M., et al. 2001, A\&A, 365, 27
  \bibitem[2007]{vKK07} van Kerkwijk, M. H., Kaplan, D. L., 2007, Ap\&SS, 308, 191 
  \bibitem[2004]{Vinkvar} Vink, J., de Vries, C. P., M\'endez, M., Verbunt, F. 2004, ApJ, 609, 75
  \bibitem[2002]{zane} Zane, S., Haberl, F., Cropper, M., et al. 2002, MNRAS, 334, 345
 
\end{thebibliography}
\end{document}